\documentclass[sn-mathphys-num]{sn-jnl}

\usepackage{graphicx}
\usepackage{multirow}
\usepackage{amsmath,amssymb,amsfonts}
\usepackage{amsthm}
\usepackage{mathrsfs}
\usepackage[title]{appendix}
\usepackage{xcolor}
\usepackage{textcomp}
\usepackage{manyfoot}
\usepackage{booktabs}
\usepackage{algorithm}
\usepackage{algorithmicx}
\usepackage{algpseudocode}
\usepackage{listings}

\usepackage[utf8]{inputenc}
\usepackage{bigints}                       
\usepackage{mathtools} 
\usepackage{epstopdf}                      
\usepackage{subcaption}                    
\usepackage{enumitem}                      
\usepackage{physics}                       
\usepackage{esdiff}                 

\renewcommand{\epsilon}{\varepsilon} 
\renewcommand{\theta}{\vartheta}     
\renewcommand{\kappa}{\varkappa}     
\renewcommand{\rho}{\varrho}         
\renewcommand{\phi}{\varphi}         
\DeclareMathOperator\sign{sgn}       

\begin{document}
\title[Article Title]{Does Bell's theorem apply if perceived pseudo-Euclidean space is emergent?}
	
\author*[1]{\fnm{Bart} \sur{Jongejan}}\email{bart.jongejan@gmail.com}
\affil*[1]{\city{Copenhagen}, \country{Denmark}}
	
\abstract{Einstein, Podolsky and Rosen (EPR) showed that it is possible to predict with certainty the value of a property \textit{without disturbing} the object in question. In contrast, Quantum Mechanics (QM) holds that if different measurement setups cannot coexist, then predictions about those can neither. Using an EPR-inspired experiment with distantly separated measurements on pairs of entangled spinning particles, Bell proved that no local hidden variable (HV) theory can describe reality in more detail than QM. However, it is possible to conceive a viable HV theory based on the assumption that the perceived structure of spacetime is emergent from a hidden curved spacetime. According to this theory, locality can be maintained for each of the measurements while what is perceived as non-locality can be ascribed to the emergence of spacetime correlations between the instruments of the two parties. The theory predicts correlations that agree with QM, provided that the hidden spacetime has three spatial dimensions. If it had fewer than three dimensions, the CHSH inequality would not be violated and if more, Tsirelson's bound would be violated. According to this HV theory, the laboratory frame of reference is a corollary of correlations of the type that are the subject of Bell's thought experiment.
}


\maketitle
	
\section{Introduction}\label{intro}
	
In a scenario envisaged by Einstein, Podolsky and Rosen \cite{PhysRev.47.777}, two parties, Alice and Bob, each can perform a measurement on one member of a pair of particles with known combined physical properties. Nothing is known about the individual properties of the particles. When Alice has measured an, until then, unknown property of her particle, she immediately knows the same property of Bob's particle. If Alice's instrument has several settings from which she can choose, she could have chosen to measure another property of her particle instead and she would have known the other property of Bob's particle. Because Alice, by measuring a property of her particle, does nothing to make Bob's particle have the corresponding property, Alice concludes that Bob's particle must have all these properties whether Alice performs measurements with her instrument or not. Bob, who has an instrument for measuring an unknown property of his particle, comes to the same conclusion regarding Alice's particle. EPR concluded that the unknown properties of the two particles are `elements of reality'. In stark contrast to EPR, the prevailing interpretation of QM holds that it is not meaningful to speak about a property except if an instrument is set up to measure it.

Since Bell's article \cite{PhysicsPhysiqueFizika.1.195} and the ensuing confirming experimental tests by Aspect \cite{PhysRevLett.49.91} and others, it seems that the EPR stance is untenable. Bell hypothesised that the unknown properties of Alice's and Bob's particles are always present and are determined by a hidden variable $\lambda$ shared between the particles before they arrive at Alice and Bob. Bell proved that no such theory can reproduce the predictions of QM. 

Alice's and Bob's instruments are supposed to be distant from each other in Bell's hypothetical shared variable theory. Henceforth, such theories will be referred to as local HV (LHV) theories because Bell uses spatial distance and a locality argument to ensure that Alice freely can choose \textit{any} setting $\hat{a}$ for her instrument without disturbing Bob's particle and thereby influence the result of Bob's measurement for which he has randomly chosen setting $\hat{b}$, and \textit{vice versa}.

Bell's theorem says that if outcomes of measurements are decided by a local hidden variable, then the expectation value $P^{LHV}\left(a,b\right)$ of the product of the outcomes ($-1$ or $1$) of Alice's and Bob's instruments will necessarily not agree with the experimentally testable value $P^{QM}\left(a,b\right)$ that is predicted by QM.

The aim of the present article is to show that Bell's theorem does not apply if one takes a strict interpretation of the foundations of general relativity into account and is ready to accept that the (pseudo)Euclidean spacetime as we perceive it is emergent from a hidden curved spacetime.
In this hidden realm reign Einstein's locality and separability, but the perceived spacetime, which has a (pseudo)Euclidean structure, may show artefacts of inseparability.
The idea that spacetime is emergent is not new.
Loop quantum gravity bases the structure of space time on spin networks \cite{Penrose_1972}, and
according to string theory, spacetime is likely an  emergent classical concept \cite{Boi2022}.

If experiments involving entanglement, such as Bell-type experiments, demonstrate the emergence of spacetime, then the story of a pair of spin component measurements must not be told as a prediction, but rather as a history, an explanation in terms of visible and hidden causes.
Since the settings $\hat{a}$ and $\hat{b}$ are known to Alice and Bob by their positions in an emergent space, we should not understand $\hat{a}$ and $\hat{b}$ as input parameters but rather as results. 

The structure of this article is as follows.
Sec. \ref{universalcor} recapitulates the expectation value of the product of two spin component measurements on a pair of particles that are known to have opposite spins. Bell's template for a LHV theory is introduced. In Sec. \ref{hicoeman}, Bell's LHV theory is reformulated by replacing the hidden variable $\hat{\lambda}$ and the settings $\hat{a}$ and $\hat{b}$ with three scalars $Z_{a\lambda}$, $Z_{b\lambda}$ and $Z_{ab}$. The same section presents an alternative theory that can reproduce the predictions of QM.
Sec. \ref{CHSHineq} discusses the CHSH inequality.
The alternative theory, similar to QM, violates the CHSH inequality, provided that space has at least three dimensions.
Moreover, Tsirelson's bound would be violated if space had more than three dimensions.
Sec. \ref{bethnoap} shows why Bell's reasoning has no effect on the presented LHV theory.
Sec. \ref{spacetimeformula} proposes a toy HV theory that suggests that spin is a consequence of a curved geometry.   This theory is a concrete instance of the alternative theory introduced in Sec. \ref{hicoeman}. The main text ends with the conclusions in Sec. \ref{conclusion}. Appendix \ref{expvalfunangsep} describes how to compute the classical correlation between two spin components. The details of the vector free formulation of Bell's and the alternative HV theories are presented in Appendix \ref{jacobian}. Appendix \ref{undicodspher} addresses the relationship between the number of spatial dimensions and the degree of violation of the CHSH inequality. Appendix \ref{geodesics} presents formulas related to the curved spacetime discussed in Sec. \ref{spacetimeformula}.

\section{Shared variables as causes of the correlations between observations of angular momentum}\label{universalcor}
	
To test EPR's conclusion, Bohm and Aharonov \cite{PhysRev.108.1070} proposed a scenario in which pairs of particles with opposite but unknown spins are separated and sent to  Alice and Bob, who each measure one spin component using instruments with settings $\hat{a}$ and $\hat{b}$. 
	
When observing a classical object with angular momentum (or spin)  $\vec{s}$ along a direction $\hat{a}$, one measures a scalar value $s_A$ that is the component of the angular momentum $\vec{s}$ in the direction $\hat{a}$.
	\begin{equation}
		\begin{aligned}\label{component}
			s_A=\hat{a}.\vec{s}
		\end{aligned}
	\end{equation}
\noindent Assuming that $\hat{s}$ varies across all spatial directions, the correlation $P\bigl(\hat{a},\hat{b}\bigr)$ between the spin component measurements is obtained by normalising the expectation value $\mathbb{E}[s_A s_B]$ of the product of the two spin components to the interval $\left[-1,1\right]$.
	\begin{equation}
		\begin{aligned}\label{expval}
			\mathbb{E}\left[s_A s_B\right]=\frac{\hat{a}.\hat{b}}{3}{\|\vec{s}\|}^2
		\end{aligned}
	\end{equation}  
	\begin{equation}
		\begin{aligned}\label{corr}
			P\bigl(\hat{a},\hat{b}\bigr) = \frac{\mathbb{E}\left[s_A s_B\right]}{\mathbb{E}\left[s_A s_A\right]} =  \hat{a}.\hat{b}
		\end{aligned}
	\end{equation}  
\noindent Eqs. \eqref{expval} and \eqref{corr} are valid for all objects that have angular momentum, irrespective of the total amount of angular momentum of the object. Classically, \eqref{expval} can be computed by integrating $\bigl(\hat{a}.\hat{s}\bigr) \bigl(\hat{b}.\hat{s}\bigr)$ over all directions $\hat{s}$, see Appendix \ref{expvalfunangsep}:
	\begin{equation}
		\begin{aligned}[b]\label{inthats}
			\mathbb{E}\left[s_A s_B\right]={\|\vec{s}\|}^2\frac{\int{\bigl(\hat{a}.\hat{s}\bigr) \bigl(\hat{b}.\hat{s}\bigr) \dd{\hat{s}}}}{\int{\dd{\hat{s}}}}
		\end{aligned}
	\end{equation}
\noindent However, according to QM, \eqref{component} and \eqref{inthats} are not correct, because spin components cannot take values from a continuous spectrum. QM does not offer an alternative to \eqref{component} that, when inserted in a formula such as \eqref{inthats}, reproduces the predictions of QM. 

In \cite{PhysicsPhysiqueFizika.1.195} Bell proposed generally formulated replacements for \eqref{component} and \eqref{inthats} and then showed that the predictions by QM could not possibly be reproduced.
To take into account the spectrum for spin-$\frac{\hbar}{2}$ particles, Bell replaced $\hat{s}$ with an unspecified shared variable $\lambda$ that travels with each of the two particles, replaced the factors $\bigl(\hat{a}.\hat{s}\bigr)$ and $\bigl(\hat{b}.\hat{s}\bigr)$ with functions $A\bigl(\hat{a},\lambda\bigr)$ and $B\bigl(\hat{b},\lambda\bigr)$ that could only take two values, $+1$ and $-1$, and included an unspecified probability measure $\rho\left(\lambda\right)$ in the integrand. The factor -1 is present because Alice's and Bob's particles have antiparallel spins: 
	\begin{equation}
		\begin{aligned}\label{Bell2}
			P^{LHV}\bigl(\hat{a},\hat{b}\bigr)=-\int \rho\left(\lambda\right) A\bigl(\hat{a},\lambda\bigr) B\bigl(\hat{b},\lambda\bigr) \dd{\lambda}
		\end{aligned}
	\end{equation} 
\noindent Bell argued that any LHV theory of the form \eqref{Bell2} obeys a particular inequality that the predictions by QM sometimes violate, thus showing that \eqref{Bell2} cannot reproduce \eqref{corr}.
	\begin{equation}
		\begin{aligned}\label{Bell2ne}
			P^{LHV}\bigl(\hat{a},\hat{b}\bigr) \ne P\bigl(\hat{a},\hat{b}\bigr)
		\end{aligned}
	\end{equation}
All discussions about LHV theories depart from Bell's abstract LHV theory \eqref{Bell2}. It is generally accepted that any HV theory that does reproduce QM cannot have the form \eqref{Bell2} and therefore is necessarily non-local. This is quite clear if the functions $A$ and $B$ were of the form $A\bigl(\hat{a},\lambda,\hat{b}\bigr)$ and $B\bigl(\hat{b},\lambda,\hat{a}\bigr)$ respectively, implying that the results of Alice's and Bob's instruments are disturbed by the setting of the other party's instrument. Also a probability density $\rho\bigl(\lambda,\hat{a},\hat{b}\bigr)$ can be regarded as breaking separability, but \cite{PhysicsPhysiqueFizika.1.195} is not explicit about this.

Bell not only defined the general form of LHV theories, but also proved that no LHV theory can reproduce QM. Therefore, presenting an LHV theory and saying that that theory reproduces QM becomes a balancing act. I keep my balance by showing that Bell's derivation from the ourset, but only implicitly, denies that spacetime relations are emergent.

\section{The hidden cost of an emergent angle}\label{hicoeman}
Bell illustrated  \eqref{Bell2} with a concrete LHV theory in which $\lambda$ is a uniformly distributed unit vector $\hat{\lambda}$, which means that the probability density $\rho\left(\lambda\right)$ is constant. The outcome of a spin component measurement in the direction $\hat{a}$ is $A\bigl(\hat{a},\hat{\lambda}\bigr)=\sign\bigl(\hat{a}.\hat{\lambda}\bigr)$. Assuming that vectors are not merely uniquely defined in reference systems local to Alice and Bob but also in a common reference system, \eqref{Bell2} becomes  
	\begin{equation}
		\begin{aligned}\label{Bell3}
			P^{\text{LHV}}\bigl(\hat{a},\hat{b}\bigr)=-\frac{ \int   \sign\bigl(\hat{a}.\hat{\lambda}\bigr) \sign\bigl(\hat{b}.\hat{\lambda}\bigr) \dd{\hat{\lambda}}}{\int \dd{\hat{\lambda}}}
		\end{aligned}
	\end{equation} 
Henceforth, in the present paper, to eliminate the minus sign in later equations, we let Alice and Bob use point reflected spatial frames of reference.

\begin{figure}[tbp]
	\begin{subfigure}[h]{0.37\linewidth}
		\includegraphics[scale=0.35]{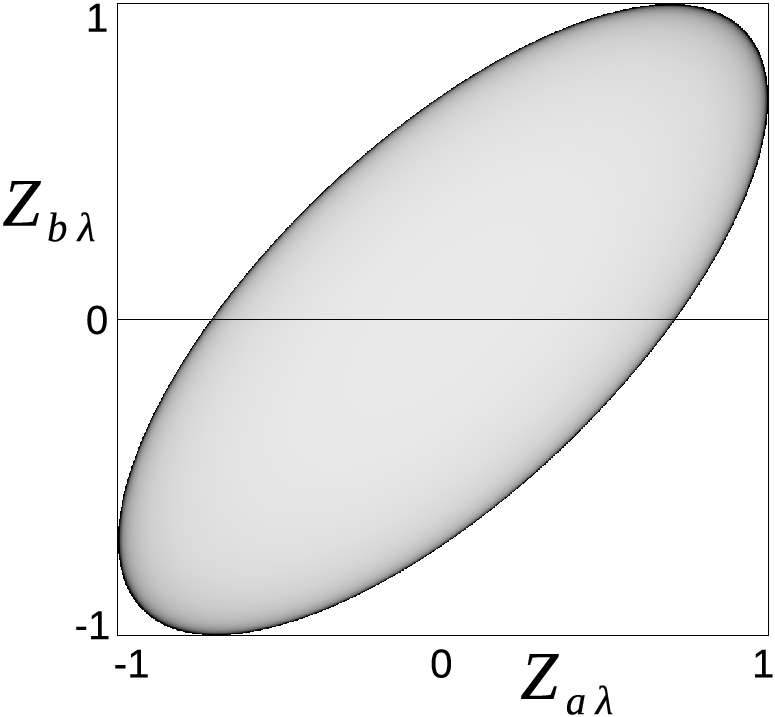}
		\caption{Bell's LHV formula}
		\label{fig:1a}
	\end{subfigure}
	\hfill
	\begin{subfigure}[h]{0.37\linewidth}
		\includegraphics[scale=0.35]{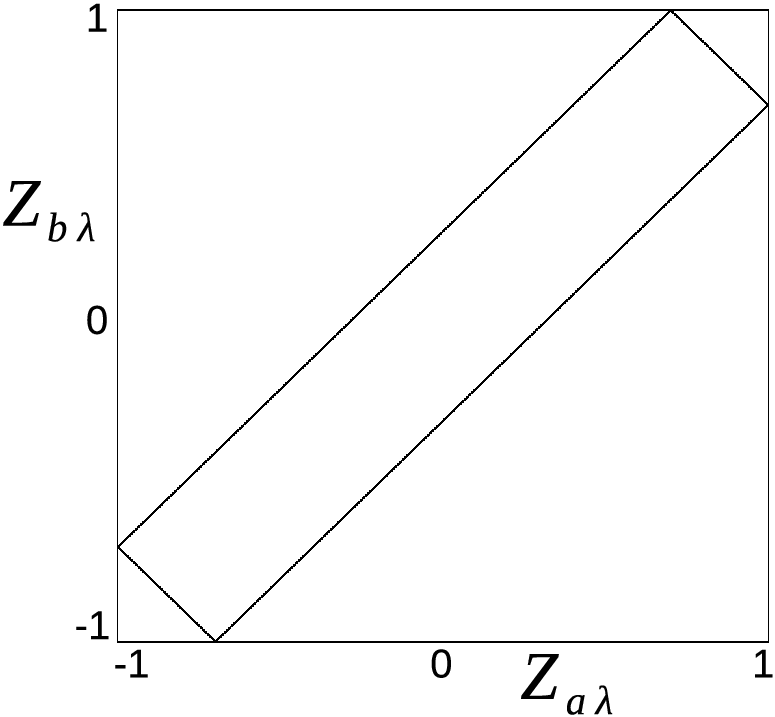}
		\caption{the alternative formula}
		\label{fig:1b}
	\end{subfigure}
	\caption{Left:Distribution of $Z_{a\lambda}$ and $Z_{b\lambda}$ for $Z_{ab}=0.7$. The distribution density is zero in the white areas. Dark = high density. Right: the allowed values of $Z_{b\lambda}$ given variable $Z_{a\lambda}$ and a fixed $Z_{ab}=0.7$.}
\end{figure}

Eq. \eqref{Bell3} does not tell how many scalar numbers are involved. It can be six if we assume that $\hat{a}$, $\hat{b}$ and $\hat{\lambda}$ each have two components. Without seriously limiting generality, we will rewrite \eqref{Bell3} using just three quantities. We do so because it will enable us to do a recount under the premise that spacetime is emerging.
First, under the conditions of Bell's experiment,  $\hat{a}.\hat{\lambda}$, $\hat{b}.\hat{\lambda}$ and $\hat{a}.\hat{b}$ vary uniformly in the interval $\left[-1,1\right]$ as we let the unit vectors $\hat{a}$, $\hat{b}$ and $\hat{\lambda}$ vary uniformly over all directions in 3-d space.  
Let
$\hat{a}.\hat{\lambda}=Z_{a\lambda}$, $\hat{b}.\hat{\lambda}=Z_{b\lambda}$ and $\hat{a}.\hat{b}=Z_{ab}$. 
Now we can eliminate $\hat{\lambda}$ from \eqref{Bell3}.
The differential $\dd{\hat{\lambda}}=\dd{\cos(\theta)}\dd{\phi}$ becomes:
	\begin{equation}
		\begin{aligned}\label{dlambd}
			\dd{\hat{\lambda}} =f_{Z_{ab}}\left(Z_{a\lambda},Z_{b\lambda}\right) \dd{Z_{b\lambda}}\dd{Z_{a\lambda}} 
		\end{aligned}
	\end{equation} 
\noindent where  $f_{Z_{ab}}\left(Z_{a\lambda},Z_{b\lambda}\right)$ is the following Jacobian determinant (see Appendix \ref{jacobian}):
	\begin{equation}
		\begin{aligned}[b]\label{dlambda}
			f_{Z_{ab}}\left(Z_{a\lambda},Z_{b\lambda}\right)= \left|\frac{\partial\left(\cos{\theta},\phi\right)}{\partial\left(Z_{a\lambda},Z_{b\lambda}\right)}\right| = {\left(8\pi\sqrt{1+2Z_{a\lambda}Z_{b\lambda}Z_{ab}-Z_{a\lambda}^2-Z_{b\lambda}^2-Z_{ab}^2}\right)}^{-1}
		\end{aligned}
	\end{equation}
\noindent We obtain the following equivalent of \eqref{Bell3}:
	\begin{equation}
		\begin{aligned}\label{Bell4}
			P^{\text{LHV}}\left(Z_{ab}\right)&=\frac{\int_{-1}^{1}\int_{l}^{h}  f_{Z_{ab}}\left(Z_{a\lambda},Z_{b\lambda}\right)  \sign\left(Z_{a\lambda}\right) \sign\left(Z_{b\lambda}\right) \dd{Z_{b\lambda}}\dd{Z_{a\lambda}}}{\int_{-1}^{1}\int_{l}^{h} f_{Z_{ab}}\left(Z_{a\lambda},Z_{b\lambda}\right) \dd{Z_{b\lambda}}\dd{Z_{a\lambda}}}\\
\text{where}\quad l&=Z_{a\lambda}Z_{ab}-\sqrt{\left(1-Z_{ab}^2\right)\left(1-Z_{a\lambda}^2\right)}\\
\text{and}\quad h&=Z_{a\lambda}Z_{ab}+\sqrt{\left(1-Z_{ab}^2\right)\left(1-Z_{a\lambda}^2\right)}
		\end{aligned}
	\end{equation} 
The integration interval for $Z_{a\lambda}$ is $\left[-1,1\right]$, but for $Z_{b\lambda}$ the boundaries $l$ and $h$ are dependent on both $Z_{ab}$ and $Z_{a\lambda}$ as a consequence of the triangle inequality \eqref{triangineq}. In other words, $Z_{a\lambda}$, $Z_{b\lambda}$ and $Z_{ab}$ are somewhat, but not completely, independent. See Fig. \ref{fig:1a} for an illustration.
\begin{equation}
	\begin{aligned}\label{triangineq}
		1+2Z_{a\lambda}Z_{b\lambda}Z_{ab}-Z_{a\lambda}^2-Z_{b\lambda}^2-Z_{ab}^2 \ge 0
	\end{aligned}
\end{equation}
Being a reformulation of \eqref{Bell3}, \eqref{Bell4} does not reproduce the QM predictions, but below I will show how the same $Z_{a\lambda}$, $Z_{b\lambda}$ and $Z_{ab}$ figure in an alternative HV theory that reproduces QM and that holds that spacetime is emergent.

If we assume that the structure of spacetime is emergent, we must assume that the angle between settings $\hat{a}$ and $\hat{b}$ is emergent. This can work as follows. Assume that there is an underlying curved spacetime, the `underworld', where the hidden variable $\hat{\lambda}$ resides. Let $\hat{\lambda}$ be a unit vector in three spatial dimensions. There is access to $\hat{\lambda}$ at both Alice's and Bob's locations. 
During each pair of measurements, Alice's actual setting $\hat{a}$ also exists in the underworld and makes an angle $\arccos{\bigl(\hat{a}.\hat{\lambda}\bigr)}$ with the hidden variable. 
If that angle is smaller than $\pi/2$, she measures $+1$; otherwise, she measures $-1$. Likewise, Bob's actual setting $\hat{b}$ also exists in the underworld and makes an angle $\arccos{\bigl(\hat{b}.\hat{\lambda}\bigr)}$ with the hidden variable. His measurement results are determined in the same way as those of Alice.
The angle between $\hat{a}$ and $\hat{b}$ is to emerge from these two angles $\arccos{\bigl(\hat{a}.\hat{\lambda}\bigr)}$ and $\arccos{\bigl(\hat{b}.\hat{\lambda}\bigr)}$. Let us for the moment just assume (I will return to this in Sec. \ref{spacetimeformula}) that there are no other variables that have a role in the emergence of the angle between $\hat{a}$ and $\hat{b}$, so there is an equation that is satisfied by the angles between $\hat{a}$, $\hat{b}$ and $\hat{\lambda}$. We can reformulate this requirement in terms of $Z_{a\lambda}$,  $Z_{b\lambda}$ and  $Z_{ab}$.
We assume, as before, that there is a function $Z_{a\lambda}$ of the angle between $\hat{a}$ and $\hat{\lambda}$ that varies uniformly in the interval $\left[-1,1\right]$ as $\hat{\lambda}$ takes all possible positions relative to $\hat{a}$ in the underworld. Similarly, we assume that $Z_{b\lambda}$ varies uniformly in $\left[-1,1\right]$ as $\hat{\lambda}$ varies uniformly over all possible directions relative to $\hat{b}$ in the underworld.
As above, we assume that  $Z_{ab}$ is defined by an equation involving  $Z_{a\lambda}$ and  $Z_{b\lambda}$ and no other variables.
Eq. \eqref{Zeq1} is such an equation. This equation ensures that $Z_{ab}$ depends linearly on uniformly and independently varying $Z_{a\lambda}$ and $Z_{b\lambda}$, and therefore also varies uniformly.
\begin{equation}
	\begin{aligned}\label{Zeq1}
&\left(1+Z_{ab}+Z_{a\lambda}+Z_{b\lambda}\right)\\
\times&\left(1-Z_{ab}-Z_{a\lambda}+Z_{b\lambda}\right)\\
\times&\left(1+Z_{ab}-Z_{a\lambda}-Z_{b\lambda}\right)\\
\times&\left(1-Z_{ab}+Z_{a\lambda}-Z_{b\lambda}\right)\\
 = 0&
	\end{aligned}
\end{equation}
\begin{figure}[tbp]
	\includegraphics[scale=0.8]{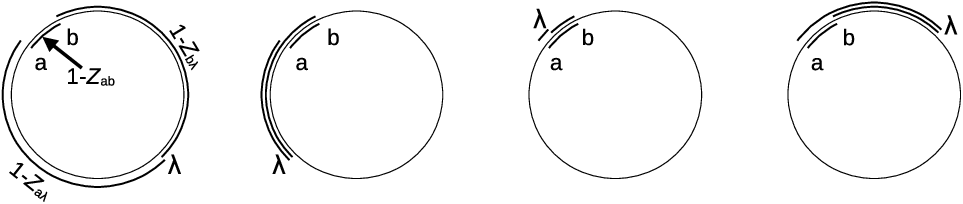}
	\caption{Integration path for $Z_{ab}=0.7$. As $\lambda$ walks along the closed path with length $4$, the points $a$, $b$ and $\lambda$ are connected by one short ($0\le length <2$) and one long ($2 \le length < 4$) arc. $Z_{xy}$ is defined as $Z_{xy}=\left(1-\text{short arc length from } x \text{ to } y\right)$  .}
	\label{fig:2}
\end{figure}
Fig. \ref{fig:1b} shows the values of $Z_{a\lambda}$ and $Z_{b\lambda}$ where \eqref{Zeq1} is obeyed, given a value of $Z_{ab}$. 
One can check that if \eqref{Zeq1} is fulfilled, then the triangle inequality \eqref{triangineq} is also satisfied.
Now, considering \eqref{Bell2}, we see that there is yet another equation to be fulfilled. In \eqref{Bell2}, $\hat{a}$ and $\hat{b}$ are not variables but parameters of the left-hand side, $P^{LHV}\bigl(\hat{a},\hat{b}\bigr)$, so $Z_{ab}$ is constant under the integral.
Given that there are three angles between three unit vectors and that there are two equations not to be violated, we conclude that only one free variable remains, not two, as in the case of Bell's illustrative LHV theory, where $\lambda$ requires two values to fix a direction in 3-d space.  
Thus, the assumption that spacetime is emergent leads to an HV theory that has a hidden variable with only one degree of freedom. In that 1-dimensional domain, $\lambda$ has a probability density $\rho\left(\lambda\right)$ that is independent of either instrument setting. This seems to be a non-local HV theory if one considers that $Z_{ab}$,  $Z_{a\lambda}$ and  $Z_{b\lambda}$ are subject to a constraint stronger than the triangle inequality, and becomes an LHV theory if one considers that the problem has fewer degrees of freedom than assumed initially.

As the free variable, we can arbitrarily choose $Z_{a\lambda}$, whereby $Z_{b\lambda}$ becomes the dependent variable. Integration over all $Z_{a\lambda}$ where \eqref{Zeq1} is obeyed gives:
\begin{equation}
	\begin{aligned}\label{HV1}
		P^{\text{ALT}}\left(Z_{ab}\right)=&\frac{\int_{-1}^{1}   \sign\left(Z_{a\lambda}\right) \sign\left(Z_{b\lambda}\left(Z_{ab},Z_{a\lambda}\right)\right)\dd{Z_{a\lambda}}}{\int_{-1}^{1} \dd{Z_{a\lambda}}} \\
		=1/4\,&\left[\ \ \int_{-1}^{-Z_{ab}}                                  \sign\left(Z_{a\lambda}\right) \sign\left(-1-Z_{ab}-Z_{a\lambda}\right)\dd{Z_{a\lambda}}\right.\\
		&\ +\int_{-Z_{ab}}^{1}\                                                 \sign\left(Z_{a\lambda}\right) \sign\left(-1+Z_{ab}+Z_{a\lambda}\right)\dd{Z_{a\lambda}}\\
		&\ +\int_{Z_{ab}}^{1}\ \ \;                                               \sign\left(Z_{a\lambda}\right) \sign\left(+1+Z_{ab}-Z_{a\lambda}\right)\dd{Z_{a\lambda}}\\
		&\left.\ +\int_{-1}^{Z_{ab}}\ \,                                        \sign\left(Z_{a\lambda}\right) \sign\left(+1-Z_{ab}+Z_{a\lambda}\right)\dd{Z_{a\lambda}}
		\right]\\
		=\ Z_{ab}&
	\end{aligned}
\end{equation}
Fig. \ref{fig:2} illustrates the integration intervals in \eqref{HV1} along each of the four sides of the slanted rectangle in Fig. \ref{fig:1b}. The first integration is along the short downwards slope in the bottom left of Fig. \ref{fig:1b}. From there the integration continues in the counterclockwise direction.

Because we define $\hat{a}.\hat{b}=Z_{ab}$ and because the QM correlation $\left\langle \hat{\sigma}.\hat{x} \  \hat{\sigma}.\hat{y}\right\rangle = \cos(\angle_{ab})$\footnote{Remember that we said that Bob's coordinate axes are mirrored with respect to Alice's.} we have, with $\angle_{ab}$ the angle between $\hat{a}$ and $\hat{b}$,
	\begin{equation}
		\begin{aligned}\label{HV2}
			P^{\text{ALT}}\left(Z_{ab}\right)=Z_{ab}=\hat{a}.\hat{b}=\cos(\angle_{ab})=\bigl\langle \hat{\sigma}.\hat{a} \  \hat{\sigma}.\hat{b}\bigr\rangle=P^{QM}\bigl(\hat{a},\hat{b}\bigr)
		\end{aligned}
	\end{equation} 
We can more clearly see that this theory is local by introducing a parameter $\lambda$ that is uniformly distributed in the interval $\left[0,4\right)$,
going around the circle in Fig. \ref{fig:2}.
The settings of the instruments are likewise numbers $a,b \in \left[0,4\right)$. Each pair of points on the circle is connected by two arcs.
The two arc lengths are defined as ${\left|xy\right|}_1 = \left|y-x\right|$ and ${\left|xy\right|}_2 = 4-{\left|xy\right|}_1$. 
Define ${\left|xy\right|}=\min\left({\left|xy\right|}_1,{\left|xy\right|}_2\right)$.
The relation between the arc lengths and the $Z$ values is $Z_{xy}=1-{\left|xy\right|}$.
A measurement by Alice is $+1$ if $Z_{a\lambda} > 0$, that is, if $\left|a\lambda\right|<1$, so $A\left(a,\lambda\right)=\sign\left(1-\left|a\lambda\right|\right)$. Likewise, $A\left(b,\lambda\right)=\sign\left(1-\left|b\lambda\right|\right)$.
The resulting formula for the expectation value $P^{\text{ALT}}\left(a,b\right)$ is in full agreement with \eqref{Bell2}:
\begin{equation}
	\begin{aligned}\label{Pparm}
		P^{\text{ALT}}\left(a,b\right)=1/4\int_{0}^{4}
		A\left(a,\lambda\right) A\left(b,\lambda\right)		\dd{\lambda}
	\end{aligned}
\end{equation}
Eq. \eqref{HV2} may seem to prove that \eqref{HV1} reproduces the predictions by QM, but that is not the case. We must consider that we could have given $Z_{ab}$ a different geometric interpretation than the cosine of the angle between $\hat{a}$ and $\hat{b}$. To determine why we choose $\hat{a}.\hat{b}=Z_{ab}$, we must examine the CHSH inequality.

\section{Shared variables and the CHSH inequality}\label{CHSHineq}
Clauser, Horne, Shimony and Holt (CHSH) \cite{PhysRevLett.23.880} considered the following formula:
\begin{equation}
	\begin{aligned}\label{Sform}
		S = 
		P\bigl(\hat{a}_1,\hat{b}_1\bigr) +P\bigl(\hat{a}_2,\hat{b}_1\bigr)+P\bigl(\hat{a}_2,\hat{b}_2\bigr) - P\bigl(\hat{a}_1,\hat{b}_2\bigr)
	\end{aligned}
\end{equation} 
\noindent CHSH showed that if the correlation value is $P\bigl(\hat{a},\hat{b}\bigr) = P^{LHV}\bigl(\hat{a},\hat{b}\bigr)$ (see \eqref{Bell2}), then the inequality \eqref{CHSHineq1} applies for any choice of $\hat{a}_1$, $\hat{a}_2$, $\hat{b}_1$ and $\hat{b}_2$.
\begin{equation}
	\begin{aligned}\label{CHSHineq1}
		\left|S\right| \le 2 
	\end{aligned}
\end{equation} 
\noindent However, for the QM correlation $P^{QM}\bigl(\hat{a},\hat{b}\bigr) = \bigl\langle \hat{\sigma}.\hat{a} \  \hat{\sigma}.\hat{b}\bigr\rangle$ (see \eqref{HV2}) the CHSH inequality \eqref{CHSHineq1} does not hold for some choices of the parameters. This finding proved, according to CHSH, that LHV theories of the form \eqref{Bell2} cannot reproduce QM. 
According to QM, the maximum value of $\left|S\right|$ is found when $\hat{a}_1$, $\hat{b}_1$, $\hat{a}_2$ and $\hat{b}_2$ are coplanar unit vectors,  $\angle_{a_1b_1} = \angle_{a_2b_1}$ = $\angle_{a_2b_2} = \frac{\pi}{4}$ and   $\angle_{a_1b_2} = \frac{3 \pi}{4}$:
	\begin{equation}
		\begin{aligned}\label{CHSHineq2}
			&\left\langle \hat{\sigma}.\hat{a}_1 \  \hat{\sigma}.\hat{b}_1\right\rangle
			+\left\langle \hat{\sigma}.\hat{a}_2 \  \hat{\sigma}.\hat{b}_1\right\rangle
			+\left\langle \hat{\sigma}.\hat{a}_2 \  \hat{\sigma}.\hat{b}_2\right\rangle
			-\left\langle \hat{\sigma}.\hat{a}_1 \  \hat{\sigma}.\hat{b}_2\right\rangle\\
			=&\cos( \frac{\pi}{4})+\cos( \frac{\pi}{4})+\cos( \frac{\pi}{4})-\cos( \frac{3 \pi}{4})=2\sqrt{2}
		\end{aligned}
	\end{equation} 
\noindent The value $2\sqrt{2}$ is Tsirelson's bound, see \cite{Cirelson1980QuantumGO}.

Sec. \ref{hicoeman} showed that there is a correlation $P^{\text{ALT}}\left(Z_{ab}\right)=Z_{ab}$ that reproduces QM if $Z_{ab}=\hat{a}.\hat{b} = \cos(\angle_{ab})$.
By defining $Z_{ab}$ in this way, we obtain a variable that is uniformly distributed over the interval $\left[-1,1\right]$ when $\hat{a}$ and $\hat{b}$ are uniformly (isotropically) distributed over all directions, but this is only true for a uniform distribution on a 2-sphere. In fewer or more dimensions, $\hat{a}.\hat{b}$ is not uniformly distributed as $\hat{a}$ and $\hat{b}$ vary isotropically.
Appendix \ref{undicodspher} shows how to compute a uniformly distributed $Z_{ab}$ for any number of spatial dimensions.
The upper bound of the expression $\left|Z_{a_1b_1}+Z_{a_2b_1}+Z_{a_2b_2}-Z_{a_1b_2}\right|$ increases with the number of dimensions. 
Tsirelson's bound is the value for three spatial dimensions.
In higher dimensions, ${\left|S\right|}_{max}$ approaches the ``superquantum'' correlation \cite{Popescu1994QuantumNA} limit of $4$; see Table \ref{tab:chshmx}.
	\begin{table}[tbp]
		\begin{tabular}{ c c c c c c}
			dimension & $\left|S\right|_{max}$ & dimension & $\left|S\right|_{max}$ & dimension & $\left|S\right|_{max}$\\
			2 & 2.000000 & 6 & 3.697653 & 10 & 3.940175 \\
			3 & 2.828427 & 7 & 3.800699 & 11 & 3.959522 \\
			4 & 3.273240 & 8 & 3.867418 & 12 & 3.972511 \\
			5 & 3.535534 & 9 & 3.911184 & 20 & 3.998648 \\
		\end{tabular}
		\caption{The maximum value of the CHSH expression in $n$-dimensional space for some values of $n$, $2 \le n \le 20$.}
		\label{tab:chshmx}
	\end{table}

\section{Why does Bell's theorem not apply?}\label{bethnoap}
We have seen in Sec. \ref{hicoeman} how an LHV theory that allows only one degree of freedom to the hidden variable is able to reproduce the predictions by QM.
In this section I assume that the conclusion in that section holds.
The question then arises as to how that can be, since Bell proved that that cannot be the case.

Bell uses three settings $\hat{a}$, $\hat{b}$ and $\hat{c}$ in his argumentation. If we accept Bell's reality definition, then $A\left(\hat{a},\hat{\lambda}\right)$, $A\left(\hat{b},\hat{\lambda}\right)$ and $A\left(\hat{c},\hat{\lambda}\right)$ are all definite at the same time. The first time this assumption is used is in the following formula, near the top of Bell's article on page 198:
\begin{equation}
	\begin{aligned}\label{Bell15}
		P\left(\hat{a},\hat{b}\right)-P\left(\hat{a},\hat{c}\right) = & -\int \dd{\lambda} \rho\left(\lambda\right) \left[A\bigl(\hat{a},\lambda\bigr) A\bigl(\hat{b},\lambda\bigr) - A\bigl(\hat{a},\lambda\bigr) A\bigl(\hat{c},\lambda\bigr)\right]\\
		= & \int \dd{\lambda} \rho\left(\lambda\right) A\bigl(\hat{a},\lambda\bigr) A\bigl(\hat{b},\lambda\bigr) \left[A\bigl(\hat{b},\lambda\bigr) A\bigl(\hat{c},\lambda\bigr) -1\right]
	\end{aligned}
\end{equation} 
As we saw in Fig. \ref{fig:2}, $\hat{a}$, $\hat{b}$ and $\hat{\lambda}$ can be represented together as points on a circuit with circumference $4$.
Let us choose a $\hat{c}$ in the same domain. As previously mentioned, the short arc length between two points corresponds to $1$ minus the cosine of the angle between the vectors they represent. Because of the one-dimensional nature of the domain, one of the arcs between $\hat{a}$, $\hat{b}$ and $\hat{c}$ is the sum of the other two arcs. Let us assume that the arc $\left|ab\right|$ between $\hat{a}$ and $\hat{b}$ is the sum of the arc $\left|ac\right|$ between $\hat{a}$, $\hat{c}$ and the arc $\left|bc\right|$ between $\hat{b}$ and $\hat{c}$. It can then be proved that $\bigl(\hat{a}-\hat{c}\bigr).\bigl(\hat{b}-\hat{c}\bigr) = 0$ for any choice of $\hat{c}$, where $\bigl(\hat{a}-\hat{c}\bigr)$ and $\bigl(\hat{b}-\hat{c}\bigr)$ are vectors in 3-space.
\begin{equation}
	\begin{aligned}\label{Bell16}
		\begin{rcases}
		\left|ab\right| = 1- \hat{a}.\hat{b} \\
		\left|bc\right| = 1- \hat{b}.\hat{c} \\
		\left|ac\right| = 1- \hat{a}.\hat{c} \\
		\left|ab\right| = \left|ac\right|+\left|bc\right| 
		\end{rcases}
		\longrightarrow
		\bigl(\hat{a}-\hat{c}\bigr).\bigl(\hat{b}-\hat{c}\bigr) = 0
	\end{aligned}
\end{equation}
This means that Bell's theorem is only proven for a subset of combinations of three settings, which are exactly the settings for which the following is true:
\begin{equation}
	\begin{aligned}\label{Bell17}
		1+P\left(\hat{b},\hat{c}\right)=\left|P\left(\hat{a},\hat{b}\right)-P\left(\hat{a},\hat{c}\right) \right|
	\end{aligned}
\end{equation} 
It is not difficult to show that no combination of settings in this subset violates the CHSH inequality, but no conclusions can be drawn from Bell's reasoning regarding any other combinations of settings.

Settings, according to the presented theory, are artefacts of processes in the curved underworld. A setting is perceived by an observer as permanent but can be explained as discrete events, not all of which happen at the same time. This, in turn, can explain why the third choice of a setting in \eqref{Bell15} may not be one of those interesting settings that, at the perceived level, is coplanar with the other two. The theory does not say that there are always discrete events that develop into all the settings one can think of. Even though it is always true that $A\bigl(\hat{a},\hat{\lambda}\bigr)A\bigl(\hat{a},\hat{\lambda}\bigr) = 1$, replacing $1$ by the product of measuring results creates an uninteresting expression that is never false but that only relates to cases where three settings are pointing at three corners of a hyperrectangle that is circumscribed by the unit (hyper)sphere and never to three points on a great circle, except if at least two of the three settings are collinear.

It may be felt as problematic that it is not possible to mention more than the two settings that QM allows in the same expression, except for a vanishing set of uninteresting settings. Was it not EPR's goal to show that unmeasured quantities also are elements of reality?
This conundrum can be solved by adopting a very strict interpretation of EPR's definition of reality: \textit{If, without in any way disturbing a system, we can predict (i.e. with probability equal to unity) the value of a physical quantity, then there exists an element of physical reality corresponding to this physical quantity}. From this definition, it does \textit{not} follow that each element of reality corresponds to a physical quantity with a value. 
It is not wrong for a realistic theory to be unable to assign values to each element of the reality that it explains. If spacetime is emergent, then elements of reality in the curved underworld may conspire to serve the quantity $\hat{a}$ to Alice and the quantity $\hat{b}$ to Bob but not a quantity $\hat{c}$, namely, if Alice and Bob perform a pair of spin component measurements with the instruments in setting $\hat{a}$ and $\hat{b}$, respectively, but that does not deny the reality of events that conspire to serve $\hat{c}$ to an observer during a spin component measurement where $\hat{c}$ is the setting and not $\hat{a}$ or $\hat{b}$. 

\section{Illustration}\label{spacetimeformula}
This section presents a toy theory that describes a spin-$\frac{\hbar}{2}$ particle as the observable consequence of a $\left(3+1\right)$-dimensional curved spacetime. It will be shown how spin component measurements are related to geodesics in this particular spacetime and how the constants of motion explain the fulfilment of Eqs. \eqref{Zeq1} and \eqref{HV1}.
This theory reproduces the predictions of QM when combined with an isotropy assumption. Appendix \ref{geodesics} presents a detailed analysis of the geodesic equations.
	
\begin{figure}[b]
		\includegraphics[scale=0.7]{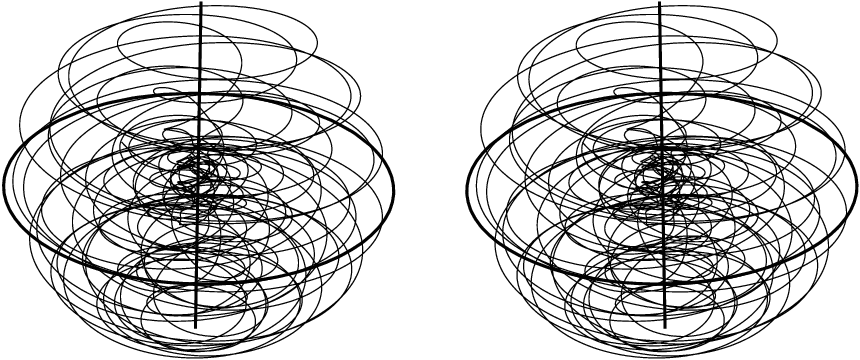}
		\caption{A bound geodesic. The plot shows the shared variable as the vertical axis and the equatorial plane as the projection of a circle. (Stereogram, left picture = left eye)}
		\label{fig:3}
	\end{figure}

\begin{figure}[tbp]
		\includegraphics[scale=0.7]{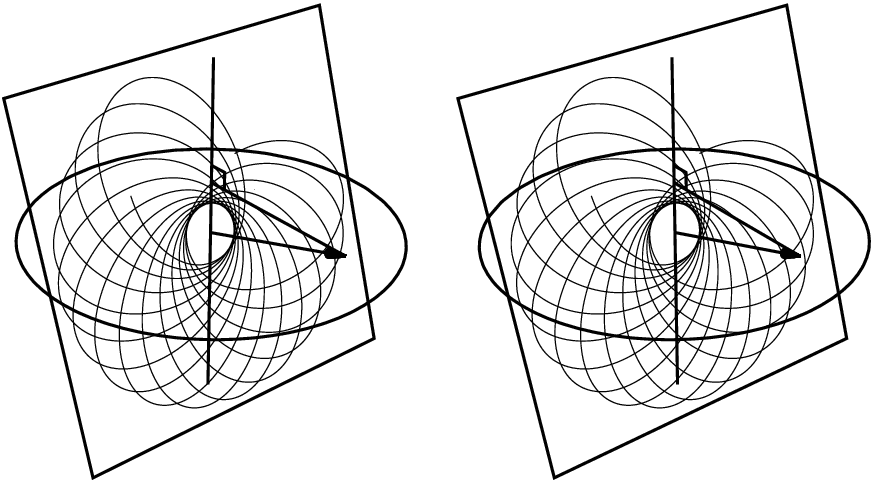}
		\caption{The same geodesic as in Fig. \ref{fig:3}, as seen by local observers along the path and at rest on spheres that are rotating to counter the fictitious force due to curvature.}
		\label{fig:4}
	\end{figure}

Consider a test particle that everywhere is subject to a fictitious force as though it is momentarily located on a rotating sphere with an equatorial speed equal to the speed of light. The picture that arises is that of an infinite set of concentric and coaxial spheres rotating with an angular velocity\footnote{Setting the velocity of light $c=1$.} of $\frac{1}{r}$, with the modification that the fictitious force is caused by the curvature of spacetime, not by rotation. The axis of symmetry is the shared variable  $\hat{\lambda}$. Fig. \ref{fig:3} depicts the spatial part of a geodesic in this spacetime. An observer at radius $r$ at rest on a rotating sphere coaxial with $\hat{\lambda}$ with angular velocity $-\frac{1}{r}$ is not subject to any acceleration. Locally, for each such observer, a passing test particle seems to follow an orbit in a slanted plane. The normal to this plane together with the sense of rotation of a test particle in the plane determine the direction in which a spin component is measured. Fig. \ref{fig:4} shows the stitched-together observations of the observers along the test particle's path.
	
A geodesic path in this spacetime has three constants of motion, together expressing the conservation of energy, the constancy of the radius of closest approach to the origin, and the inclination $\iota$ of a test particle where its polar velocity $U^{\theta}$ is zero. A positive $\iota$ corresponds to measuring spin `up' and a negative $\iota$ corresponds to spin `down'.
The inclination of the orbital plane is also the angle between the central axis and the normal to the orbital plane. 
\noindent The geodesics in this spacetime have much in common with the orbits of particles under the influence of a central force field. There are bound (captive) geodesics as well as ``attractive'' and ``repelling'' unbound (escape) geodesics, corresponding to elliptical, parabolic and hyperbolic trajectories in a central force field. The attractive and repelling unbound geodesics are analogous to the possible paths of a spin-$\frac{\hbar}{2}$ particle through a Stern Gerlach instrument, or rather, the possible paths of the instrument from the perspective of the particle.
	
There is a crucial difference between a geodesic in the proposed spacetime and a trajectory in a central force field. The latter has two scalar constants of motion to define the orbital plane. The full determination of the orientation of an orbital plane relative to another such plane again requires two constants. The orbital plane in the conjectured spacetime, on the other hand, requires only one constant of motion: the angle between the normal and the central axis. The difference in orientation between two such planes is therefore also just one constant.
	
Why is this spacetime a candidate LHV theory?
For the sake of the argument, let us assume that in the course of a large number of spin component measurements performed by Alice (Bob), $\cos(\iota_{a\left(b\right)}) = Z_{a\left(b\right)\lambda}$ is uniformly distributed in the interval $\left[-1,1\right]$, in line with a uniform (isotropic) distribution of unit vectors in flat 3-dimensional space. If this assumption holds, than also linear combinations of $Z_{a\lambda}$ and $Z_{b\lambda}$ are uniformly distributed, in particular the scalar $Z_{ab}$, $-1 \le Z_{ab} \le 1$: 
\begin{equation}
	\begin{aligned}\label{Zabdef}
		Z_{ab}= \begin{cases}
			-1-Z_{a\lambda}-Z_{b\lambda} & \text{provided that $0 \le -Z_{a\lambda} -Z_{b\lambda} \le 2$ or} \\
			+1-Z_{a\lambda}+Z_{b\lambda} & \text{provided that $0 \le \ \ \,Z_{a\lambda} -Z_{b\lambda} \le 2$ or}\\
			-1+Z_{a\lambda}+Z_{b\lambda} & \text{provided that $0 \le \ \ \,Z_{a\lambda} +Z_{b\lambda} \le 2$ or}\\ 
			+1+Z_{a\lambda}-Z_{b\lambda} & \text{provided that $0 \le -Z_{a\lambda} +Z_{b\lambda} \le 2$ or} 
		\end{cases}
	\end{aligned}
\end{equation} 
For each actual pair of settings $\left(a_m,b_n\right)$ there are two pairs $\left(Z_{a_m\lambda_1},Z_{b_n\lambda_1}\right)$ and $\left(Z_{a_m\lambda_2},Z_{b_n\lambda_2}\right)$ such that two of the four alternatives in Eq. \eqref{Zabdef} apply, thereby satisfying \eqref{Zeq1} and \eqref{HV1}.
If $Z_{ab}$ is uniformly distributed over the interval $\left[-1,1\right]$ as $a$ and $b$ are chosen at randon by Alice and Bob, then $\arccos\left(Z_{ab}\right)$ is always in agreement with the Euclidean angle $\angle_{ab}$ separating $\hat{a}$ and $\hat{b}$, confirming that $Z_{ab}$ is identical to \eqref{HV2}. This concludes the proof that a shared cause can explain the predictions of QM without the instrument setting of one party disturbing the outcome of the measurement of the other party.
	
\section{Conclusion}\label{conclusion}

The reasoning in this article is based on the assumption that the (pseudo)Euclidean spacetime we perceive is emergent from an underlying, mostly hidden spacetime that has a curved structure.
Experiments involving entanglement show how this emergence works: they enable us to operationally define geometrical relations over large spatial distances, bypassing classical means such as measuring rods, clocks and gyroscopes.
Measuring the angle between settings of spatially well separated instruments can be done in more practical ways than by repeating pairs of spin component measurements many times.
In a general relativistic context, however, it is contingent that different operational definitions of the same spatial relation are in agreement with each other; therefore, no operational definitions are redundant.

I propose a hidden variable theory that describes how a Bell-type experiment evolves in hidden, curved spacetime.

The proposed metric leads to geodesics that in some ways resemble those in a central force field, but in contrast to the latter, the geodesics in the proposed theory are not confined to orbital planes with a normal fixed in space. Instead, the axis of symmetry (the hidden `variable') and the normal to the orbital plane remain at a constant distance from each other. The angular distance between the vectors $\hat{a}$ and $\hat{b}$ in any particular pair of spin component measurements emerges from a relation involving only the inclinations of the geodesics that are followed by the measuring instruments. 

Each pair of outcomes contributes one bit of information to our perception of the angle between the settings of Alice's and Bob's instruments, and this angle, by design of the experiment, is non-local. On the other hand, the measurement of a spin component by one party is not disturbed by the setting of the instrument of the other party and is therefore free of non-locality.

Adopting an emergent flat spacetime instead of a preexisting flat space time means that we have to reckon with fewer free variables in the setup of a pair of spin component measurements, because the angle between the settings of the instruments becomes a dependent variable. This reduction can be interpreted either as more strict conditional distributions of the two settings and the hidden variable or as an event taking place in a (sub)space of fewer dimensions. In the first case it is appropriate to call the resulting theory non-local, but under the second interpretation locality survives.
 
An unintended but interesting side effect of the presented theory is that it explains why the maximum rate of violation of the CHSH inequality is equal to Tsirelson's bound. The violation would be larger if the underlying spacetime had more than three spatial dimensions, and no violation would occur if the underlying spacetime had fewer spatial dimensions.

\section*{Declarations}

	No funds, grants, or other support was received.

\noindent The author has no relevant financial or non-financial interests to disclose.	
\begin{appendices}

\section{The expectation value of the product of two spin component measurements as a function of angular separation}\label{expvalfunangsep}
	
In the case of a spin-$\frac{\hbar}{2}$ particle, with total spin angular momentum $S$, QM predicts an expectation value:
	\begin{equation}\label{quantumcase}
		\begin{aligned}[b]
			\mathbb{E}\left[s_A s_B\right]=\hat{a}.\hat{b}\left(\frac{\hbar}{2}\right)^{2}= \frac{\hat{a}.\hat{b}}{3} \left(\sqrt{\frac{1}{2}\left(\frac{1}{2}+1\right)}{\hbar}\right)^{2}= \frac{\hat{a}.\hat{b}}{3}S_{\frac{\hbar}{2}}^2
		\end{aligned}
	\end{equation}
\noindent Classically, the result of measuring a single component of the spin $\vec{s}$ can take any value between $-|\vec{s}|$ and $|\vec{s}|$. This value varies as the cosine of the angle between the rotation axis of the spinning particle and the direction of observation.
	
In the following, $\hat{a}$ and $\hat{b}$ are the two  directions of observation and $\hat{s}$ denotes the spin axis of the particle.
	\begin{equation}
		\begin{aligned}[b]\label{classicalcase}
			\hat{a}&=[1,0,0], \hat{b}=[\cos\gamma,\sin\gamma,0], \hat{s}=[\sin\theta\cos\phi, \sin\theta\sin\phi, \cos\theta]\\ 
			\hat{a}.\hat{s} &= \sin\theta\cos\phi\\
			\hat{b}.\hat{s} &= \cos\gamma\sin\theta \cos\phi+\sin\gamma\sin\theta\sin\phi\\
		\end{aligned}
	\end{equation}
	\begin{equation}
		\begin{aligned}[b]
			\mathbb{E}\left[s_A s_B\right]=&{\|\vec{s}\|}^2\frac{\int{\bigl(\hat{a}.\hat{s}\bigr) \bigl(\hat{b}.\hat{s}\bigr) \dd{\hat{s}}}}{\int{\dd{\hat{s}}}}\\
			=&\frac{{\|\vec{s}\|}^2}{4 \pi}\int\limits_0^{2 \pi}\int\limits_0^\pi\sin\theta\cos\phi\left(\cos\gamma\sin\theta \cos\phi+\sin\gamma\sin\theta\sin\phi\right) \sin \theta \ \dd\theta \ \dd\phi
		\end{aligned}
	\end{equation}
\noindent The second term vanishes because
	\begin{equation}
		\begin{aligned}
			\int\limits_0^{2 \pi}\cos\phi\sin\phi \ \dd\phi = \int\limits_0^0\sin\phi \ \dd\sin\phi = 0
		\end{aligned}	
	\end{equation}
\noindent In the first term
	\begin{equation}
		\begin{aligned}
			\int\limits_0^\pi\sin^{2}\theta \sin \theta \ \dd\theta = \int\limits_{-1}^{1} \left(1-\cos^{2}\theta\right) \ \dd\cos\theta = 2-\frac{2}{3} = \frac{4}{3}
		\end{aligned}	
	\end{equation}
\noindent and
	\begin{equation}
		\begin{aligned}[b]
			\int\limits_0^{2 \pi}\cos^{2}\phi \ \dd\phi = \pi
		\end{aligned}
	\end{equation}
\noindent Thus
	\begin{equation}
		\begin{aligned}[b]
			\mathbb{E}\left[s_A s_B\right] = \frac{{\|\vec{s}\|}^2}{4 \pi}\int \bigl(\hat{a}.\hat{s}\bigr)\bigl(\hat{b}.\hat{s}\bigr)\dd{ \hat{s}} = \frac{{\|\vec{s}\|}^2}{4 \pi}\frac{4 \pi}{3}\cos\gamma
			=  \frac{\hat{a}.\hat{b}}{3}{\|\vec{s}\|}^2
		\end{aligned}
	\end{equation}
\section{Bell's LHV model without vectors}\label{jacobian}
Our goal is to eliminate coordinates from the following formula and to use only the relations $Z_{xy}$ between $\hat{\lambda}$, $\hat{a}$, and $\hat{b}$.
	\begin{equation}
		\begin{aligned}\label{Bell3bis1}
			P^{LHV}\left(\hat{a},\hat{b}\right)=constant \int  \sign\left(\hat{a}.\hat{\lambda}\right) \sign\left(\hat{b}.\hat{\lambda}\right) \dd{\hat{\lambda}}
		\end{aligned}
	\end{equation} 
For each pair of measurements of Alice and Bob we choose a Euclidean coordinate system such that $\hat{a}$ defines the $z$-axis and $\hat{b}$ lies in the $x$-$z$-plane with a non-negative $x$-coordinate. The orientation of the $y$-axis is not important for the following. The shared variable $\lambda$ varies across all directions over a series of measurement pairs. Additionally, Alice and Bob move their instruments to arbitrary but identifiable settings between measurements so that, in the long run, the direction of their instruments is uniformly distributed over all directions on a 2-sphere. Set $\hat{\lambda} = \left[\sin{\theta}\cos{\phi},\sin{\theta}\sin{\phi},\cos{\theta}\right]$, $\hat{a}=\left[0,0,1\right]$, and $\hat{b}=\left[\sin{\beta},0,\cos{\beta}\right]$ with $0 \le \theta \le \pi$, $0 \le \phi \le 2\pi$ and $0 \le \beta \le \pi$.
	\begin{equation}
		\begin{aligned}\label{Bell3bis2}
			Z_{a\lambda} &= \hat{a}.\hat{\lambda}=\cos\theta\\
			Z_{b\lambda} &=\hat{b}.\hat{\lambda} =\sin\beta \sin\theta \cos\phi+\cos\beta \cos\theta\\
			Z_{ab} &=\hat{a}.\hat{b}= \cos\beta
		\end{aligned}
	\end{equation} 
\noindent We have, with $C$ a normalising constant,
	\begin{equation}
		\begin{aligned}\label{Bell3bis3}
			P^{LHV}\left(\hat{a},\hat{b}\right)=\ \ \ &C \int \sign\left(Z_{a\lambda}\right) \sign\left(Z_{b\lambda}\right) \sin{\theta}\ \dd{\theta} \dd{\phi}\\
			= - &C \int \sign\left(Z_{a\lambda}\right) \sign\left(Z_{b\lambda}\right) \dd{\cos{\theta}}\dd{\phi}
		\end{aligned}
	\end{equation} 
To eliminate $\cos{\theta}$ and $\phi$, we note that 
	\begin{equation}
		\begin{aligned}[b]\label{Zal}
			\dd{Z_{b\lambda}} \dd {Z_{a\lambda}} = J\left(\cos{\theta},\phi\right) \dd{\cos{\theta}}\dd{\phi} 
		\end{aligned}
	\end{equation}
\noindent where the Jacobian $J\left(\cos{\theta},\phi\right)$ is the determinant of the matrix
	\begin{equation}
		\begin{aligned}[b]\label{Za2}
			J\left(\cos{\theta},\phi\right) =& \left|\frac{\partial\left(Z_{a\lambda},Z_{b\lambda}\right)}{\partial\left(\cos{\theta},\phi\right)}\right| = \frac{\partial Z_{a\lambda}}{\partial \cos\theta}\frac{\partial Z_{b\lambda}}{\partial \phi} - \frac{\partial Z_{a\lambda}}{\partial \phi}\frac{\partial Z_{b\lambda}}{\partial \cos\theta}\\
			=& - 1.\sin\beta \sin\theta \sin\phi - 0.\sin\beta \cos\theta \sin\phi\\
			=& - \sqrt{\left(1-\cos^2\beta\right)\left(1-\cos^2\theta\right)} \sin\phi
		\end{aligned}
	\end{equation}
\noindent We derive $\sin\phi$ by eliminating the goniometric functions of $\beta$ and $\theta$ from the formula for $Z_{b\lambda}$ in \eqref{Bell3bis2}:
	\begin{subequations}
		\begin{align}
			Z_{b\lambda} &= {\left[\left(1-Z^2_{ab}\right)\left(1-Z^2_{a\lambda}\right)\right]}^{1/2} \cos\phi+Z_{ab} Z_{a\lambda}\\
			\cos\phi&=\frac{Z_{b\lambda}-Z_{a\lambda}Z_{ab}}{{\left[\left(1-Z^2_{ab}\right)\left(1-Z^2_{a\lambda}\right)\right]}^{1/2}}\\
			\sin\phi&=\sqrt{1-\cos^2\phi} = \sqrt{1-\frac{{\left(Z_{b\lambda}-Z_{a\lambda}Z_{ab}\right)}^{2}}{{\left(1-Z^2_{ab}\right)\left(1-Z^2_{a\lambda}\right)}}}\nonumber\\
			&=\sqrt{\frac{1-Z^2_{a\lambda} - Z^2_{b\lambda} - Z^2_{ab} + 2 Z_{a\lambda} Z_{b\lambda} Z_{ab} }{\left(1-Z^2_{ab}\right)\left(1-Z^2_{a\lambda}\right)}}\label{Bell3bis4}
		\end{align}
	\end{subequations} 
\noindent By inserting \eqref{Bell3bis4} into \eqref{Za2} we obtain
	\begin{equation}
		\begin{aligned}[b]\label{Za3}
			J\left(\cos{\theta},\phi\right) &= - \sqrt{\left(1-\cos^2\beta\right)\left(1-\cos^2\theta\right)} \sqrt{\frac{1-Z^2_{a\lambda} - Z^2_{b\lambda} - Z^2_{ab} + 2 Z_{a\lambda} Z_{b\lambda} Z_{ab} }{\left(1-Z^2_{ab}\right)\left(1-Z^2_{a\lambda}\right)}}\\
			&= - \sqrt{\left(1-Z^2_{ab}\right)\left(1-Z^2_{a\lambda}\right)} \sqrt{\frac{1-Z^2_{a\lambda} - Z^2_{b\lambda} - Z^2_{ab} + 2 Z_{a\lambda} Z_{b\lambda} Z_{ab} }{\left(1-Z^2_{ab}\right)\left(1-Z^2_{a\lambda}\right)}}\\
			&= - \sqrt{1-Z^2_{a\lambda} - Z^2_{b\lambda} - Z^2_{ab} + 2 Z_{a\lambda} Z_{b\lambda} Z_{ab} }
		\end{aligned}
	\end{equation}
\noindent Thus we find that 
	\begin{equation}
		\begin{aligned}[b]\label{Jac}
			J\left(Z_{a\lambda},Z_{b\lambda}\right) = J^{-1}\left(\cos{\theta},\phi\right)
			= - {\left(\sqrt{1-Z^2_{a\lambda} - Z^2_{b\lambda} - Z^2_{ab} + 2 Z_{a\lambda} Z_{b\lambda} Z_{ab}} \right)}^{-1}
		\end{aligned}
	\end{equation}
\noindent And finally
	\begin{equation}
		\begin{aligned}\label{Bell3bis1bis}
			P^{LHV}_{ab}=C \int \frac{\sign\left(Z_{a\lambda}\right) \sign\left(Z_{b\lambda}\right)}{ {\sqrt{1-Z^2_{a\lambda} - Z^2_{b\lambda} - Z^2_{ab} + 2 Z_{a\lambda} Z_{b\lambda} Z_{ab} }}} \dd{Z_{b\lambda}}\dd{Z_{a\lambda}}
		\end{aligned}
	\end{equation} 
	
\section{Uniformly distributed correlation on the \textit{d}-sphere as function of angle}\label{undicodspher}
	
The area $A$ of the hyperspherical cap of radius $r$ and height $h$ of a $d$-sphere is
	\begin{equation}
		\begin{aligned}[b]\label{hypersphericalcap}
			A_d = C \int_{0}^{\gamma} \sin[d-1](\theta) \ \dd{\theta}
		\end{aligned}
	\end{equation}
\noindent where $\gamma = \arccos(\frac{r-h}{r})$. Using integration by parts, we have, for $0 \le \gamma \le \pi$ and $d\ge 1$:
	\begin{equation}
		\begin{aligned}[b]\label{integral}
			I\left(\gamma,d \right) =\int_{0}^{\gamma} \sin[d-1](\theta) \dd{\theta}= \begin{cases}
				\gamma & \text{if $d=1$}\\
				1 - \cos{\gamma}  & \text{if $d= 2$}\\
				-\frac{1}{d-1} \sin^{d-2}{\gamma} \cos{\gamma} + \frac{d - 2}{d -1} I\left(\gamma, d - 2\right) & \text{if $d\ge 3$}
			\end{cases}
		\end{aligned}
	\end{equation}
\noindent Eq. \eqref{integral} computes a number that is proportional to the area of the $d$-hyperspherical cap that contains all points that are less than an angle $\gamma$ from the point $\gamma=0$. If $N$ points are uniformly distributed over the $d$-sphere, then a fraction $\frac{I\left(\gamma,d \right)}{I\left(\pi,d \right)} N$ is expected to lie within the spherical cap. We can formulate as follows a correlation $P\left(\gamma,d \right), -1 \le P\left(\gamma,d \right) \le 1$ that is uniformly distributed if $\gamma$ is the angle between two vectors $\hat{x}$ and $\hat{y}$ that both are chosen randomly from a uniform distribution over a $d$-sphere:
	\begin{equation}
		\begin{aligned}\label{expectationValueA}
			P\left(\gamma,d \right) = 
			1 - \frac{2 I\left(\gamma,d\right)}{I\left(\pi, d\right)}		 
		\end{aligned}
	\end{equation}
In the case of a circle ($d = 1$), $P$ varies linearly with the length $\gamma$ of the arc separating the vectors $\hat{x}$ and $\hat{y}$. In the case of a 2-sphere ($d=2$), $P$ varies as $\cos(\gamma)$

\noindent The expression
	\begin{equation}
		\begin{aligned}\label{chshmax}
			3*P\left(\gamma,d \right) - P\left(3 \gamma,d \right)		 
		\end{aligned}
	\end{equation}
\noindent reaches its maximum for $\gamma = \frac{\pi}{4}$ for all values $d > 1$ because
	\begin{equation}
		\begin{aligned}\label{chshma}
			&\diff{}{\gamma}\left[3*P\left(\gamma,n \right) - P\left(3 \gamma,d \right)\right]
			\propto  \diff{}{\gamma}\left[3*I\left(\gamma,d \right) - I\left(3 \gamma,d \right)\right]\\
			\propto &\ 3 \sin[d-1](\gamma) - \diff{3\gamma}{\gamma}\sin[d-1](3\gamma)
			\propto  \sin[d-1](\gamma) - \sin[d-1](3\gamma)
		\end{aligned}
	\end{equation}
and
	\begin{equation}
		\begin{aligned}\label{chshm}
			\sin[d-1](\frac{\pi}{4}) - \sin[d-1](\frac{3\pi}{4}) = \left(\sqrt{\frac{1}{2}}\right)^{d-1} - \left(\sqrt{\frac{1}{2}}\right)^{d-1} = 0
		\end{aligned}
	\end{equation}
which means that \eqref{chshmax} reaches an extremum for $\gamma = \frac{\pi}{4}$.
\section{Geodesics in toy spacetime}\label{geodesics}
Consider a metric $g_{ik}$
	\begin{equation}
		\begin{aligned}
			ds^2=\cos^2\vartheta dt^2 - dr^2 - r^2d\vartheta^2 - r^2\sin^2\vartheta d\varphi^2 +2r\sin^2\vartheta d\varphi dt
		\end{aligned}
	\end{equation}
To investigate how free test-particles move in this spacetime, we have to solve the equations of motion
	\begin{equation}
		\begin{aligned}
			\diff[2]{x_i}{s} =-\Gamma^i_{jk}\diff{x_j}{s}\diff{x_k}{s}
		\end{aligned}
	\end{equation}
With the shorthand notation $U^r=r'=\diff{r}{s}$, $U^{r'}=r''=\diff[2]{r}{s}$ etc., the geodesic equations are
	\begin{subequations}
		\begin{align}
			U^{t'}=&\frac{-\,U^r\sin\vartheta\left[\sin\vartheta\left(U^t-rU^\varphi\right)\right]}{r}\label{eq:ged1}\\
			U^{r'}=&\frac{\left(rU^\vartheta\right)^2-\left(rU^\varphi\sin\vartheta\right)[\sin\vartheta\left(U^t-rU^\varphi\right)]}{r}\label{eq:ged2}\\
			U^{\vartheta'}=&\frac{-\,2\,U^r\left(rU^\vartheta\right)+\frac{\cos\vartheta}{\sin\vartheta}[\sin\vartheta(U^t-rU^\varphi)]^2}{r^2}\label{eq:ged3}\\
			U^{\varphi'}=&\left.\frac{1}{ r^2\sin\vartheta}\right\{-U^r\left(rU^\varphi\sin\vartheta\right)\nonumber\\
			&\quad\quad\quad\quad\ +U^r\cos^2\vartheta\left[\sin\vartheta\left(U^t-rU^\varphi\right)\right]\nonumber\\
			&\quad\quad\quad\quad\ +\left.2\frac{\cos\vartheta}{\sin\vartheta}\left(rU^\vartheta\right)\left[\sin\vartheta\left(U^t-rU^\varphi\right)\right]\right\}\label{eq:ged4}
		\end{align}
	\end{subequations}
These equations have the following solutions:
	\begin{subequations}
		\begin{align}
			U^t&=P+\frac{X}{r}\label{eq:sol1}\\
			U^\varphi&=\frac{P-\frac{X{\rm cotg}^2\vartheta}{r}}{r}\label{eq:sol2}\\
			{\left(U^\vartheta\right)}^2&=\frac{A-\frac{X^2}{\sin^2\vartheta}}{r^4}\label{eq:sol3}\\
			{\left(U^r\right)}^2&=-\,\frac{A-X^2}{r^2}+\frac{2PX}{r}+P^2-W\label{eq:sol4}
		\end{align}
	\end{subequations}
where $P,X,A$ and $W$ are constants:
	\begin{subequations}
		\begin{align}
			&P=\cos^2\vartheta U^t+r\sin^2\vartheta U^\varphi\label{P}\\
			&X=r\sin^2\vartheta\left(U^t-rU^\varphi\right)\label{eq:gat2}\\
			&A={\left(r^2U^\vartheta\right)}^2+{\left[r\sin\vartheta\left(U^t-rU^\varphi\right)\right]}^2\label{eq:gat3}\\
			&W={\left(\cos\vartheta U^t\right)}^2-{\left(U^r\right)}^2 -{\left(rU^\vartheta\right)}^2-{\left(r\sin\vartheta U^\varphi\right)}^2+2r\sin^2\vartheta U^tU^\varphi\label{eq:gat4}
		\end{align}
	\end{subequations}
$W$ is $g_{ij} U^{i} U^{j}$, the length of the vector $U$ squared. Time-like geodesics have $W>0$, space-like geodesics have $W<0$ and light-like geodesics have $W=0$. We will restrict $W$ to the values -1, 0 and 1. This restriction removes an arbitrary scaling factor.
	
$A,P,X$ and $W$ are real numbers that can be chosen freely only to some degree. From \eqref{eq:sol3} follows that $\left|X\right|\leq\sqrt A$ and that geodesics are restricted to points where $\sin\vartheta\geq{\left|X\right|/\sqrt A}$. Geodesics with values of $\left|X\right|$ close to $\sqrt A$ stay close to the equatorial plane $\vartheta =\frac{\pi}{2}$, whereas a value of $\left|X\right|/\sqrt A$ close to zero allows the geodesic to approach the poles closely. We define $Z$ as follows:
	\begin{equation}\label{Z}
		\begin{aligned}
			Z=\frac{X}{\sqrt A}\qquad\left(-1\leq Z\leq1\right)
		\end{aligned}
	\end{equation}
The maximum inclination $\iota_{max}$ from the equatorial plane is reached when $U^\vartheta=0$, which is when $\sin\vartheta=\left|Z\right|$. Then $\iota_{max}=\arccos(Z)$.
	
We can learn much about the orbital shape and size from investigating the radial velocity. If $A\ne X^2$, then ${\left(U^r\right)}^2$ is a second order function of $1/r$, otherwise it is a first order function of $1/r$. We will now look at the constraints on $P,X,W$ and $A$ that ensure that 
	\begin{equation}
		\begin{aligned}
			{\left(U^r\right)}^2=-\,\frac{A-X^2}{r^2}+\frac{2PX}{r}+P^2-W=0
		\end{aligned}
	\end{equation}
	has positive solutions of $r$. We have:
	\begin{equation}\label{1_r}
		\begin{aligned}
			\left.\frac{1}{r}\right|_{U^r=0}=\frac{PX\pm\sqrt{AP^2-AW+X^2W}}{A-X^2}
		\end{aligned}
	\end{equation}
	The condition that there be solutions is that
	\begin{equation}
		\begin{aligned}
			P^2\geq W\left(1-\frac{X^2}{A}\right) = W\left(1-Z^2\right)
		\end{aligned}
	\end{equation}
The factor in parentheses is non-negative because $1\geq |Z|$.  If $W\leq0$, then the relation is fulfilled for all values of $P$. For positive $W$ the above relation puts a lower bound on the absolute value of $P$.
	
Because the radial parameter $r$ is non-negative, we are only interested in non-negative solutions of $1/r$. If $P > 1$, there is one positive solution, the trajectory is unbound, and the trajectory is {\tt"}hyperbolic{\tt"}. If $P=1$, then the trajectory is barely unbound and has the status of a parabolic trajectory in a central force field. If $P<1$ and $Z>0$, then there are two positive solutions, and the trajectory is bound, similar to elliptic trajectories. If $P^2=1-Z^2$ then the trajectory has a constant radius, i.e., it is comparable with circular orbits.
	
The analogy between classical orbits in central force fields and geodesics in our spacetime formula is concisely expressed by 
	\begin{equation}\label{P2_W}
		\begin{aligned}
			\frac{P^2-W}{2}=\frac{{\left(U^r\right)}^2+{\left(U^\perp\right)}^2}{2}-\frac{XP}{r}-\frac{X^2}{{2r}^2}
		\end{aligned}
	\end{equation}
	where
	\begin{equation}
		\begin{aligned}[t]
			{\left(U^\perp\right)}^2={\left(rU^\vartheta\right)}^2+\sin^2\vartheta(U^t-rU^\varphi)^2&
		\end{aligned}
	\end{equation}
Eq. \eqref{P2_W} has the signature of energy per mass, with a kinetic contribution depending on the squares of the radial and tangential velocities and a potential contribution in two parts: a long range $1/r$ attractive or repelling potential and a short range attractive $1/r^2$ potential that adds a precession of the pericentrum to the movement. The two independent constants, $P$ and $X$, together define the shape and size of an orbit.
	
\begin{figure}[tbp]
	\begin{subfigure}[h]{0.4\linewidth}
		\includegraphics[scale=0.6]{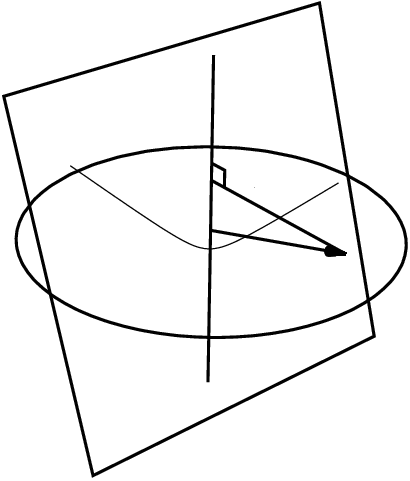}
\caption{attractive, $Z = 0.3$}
\label{fig:attractive}
	\end{subfigure}
	\hfill
	\begin{subfigure}[h]{0.4\linewidth}
		\includegraphics[scale=0.6]{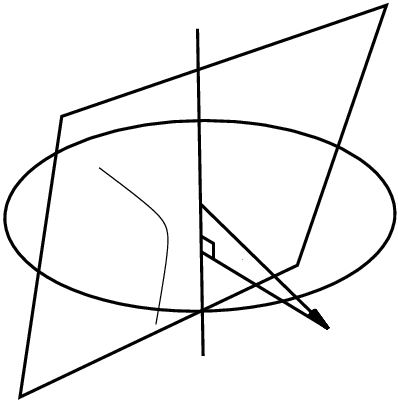}
\caption{repulsive, $Z = -0.3$}
\label{fig:repulsive}
	\end{subfigure}
	\caption{Unbound geodesics}
\end{figure}

Figures \ref{fig:attractive}-\ref{fig:repulsive} depict two unbound geodesics, differing only in the sign of $Z$ (or, equivalently, of the $X$-parameter), from the stitched-together perspectives of all observers moving with azimuthal speed $-1/r$ that are passed by the test particle. In this view, a test particle seems to move in an orbital plane.
	
The direction in which a spinning particle is deflected in a Stern-Gerlach instrument depends on the direction of the force
	\begin{equation}
		\begin{aligned}
			F_z=-\nabla\left(-\mu.B\right)=\mu_z\frac{\partial B_z\,}{\partial z}
		\end{aligned}
	\end{equation}
where $\mu$ is the magnetic moment of the particle associated with the spin of the particle and $B$ is the magnetic field of the magnet, having a gradient $\partial B_z/\partial z$. As seen from a spin-$\frac{\hbar}{2}$ particle's point of view, depending on the sign of $\mu_z$, the pole where the magnetic field is strongest is either repelled from or attracted to the particle as the particle passes through the diverging field: the two situations are not symmetric. Nor are the (unbound) geodesics of two test particles that differ only by the sign of $XP$, see Figs. \ref{fig:attractive} and \ref{fig:repulsive}. According to \eqref{P2_W}, the $1/r$ potential gives rise to either a repulsive or an attractive force, depending on the sign of $XP$. This analogy is why we can surmise that the sign of $X$ corresponds to the outcome of a spin component measurement. ($P$'s sign defines the direction of time; see \eqref{P}) Thus, according to \eqref{Z}, the outcome of a spin component measurement depends on whether the orbital plane tilts less or more than $\frac{\pi}{2}$ with respect to the equatorial plane. 
\end{appendices}

\end{document}